\input ./style/arxiv-general.cfg
\documentclass[aoas,MSNbibl,nameyear,rotating,dvips]{arximspdf}
\makeatletter
   \@ifpackageloaded{graphicx}{}{\usepackage{graphicx}}
\makeatother
\usepackage{multirow}

\doi{10.1214/15-AOAS827}
\volume{9}
\issue{3}
\pubyear{2015}
\firstpage{1247}
\lastpage{1277}
\docsubty{FLA}

\makeatletter

\newcommand{\eqref}[1]{(\ref{#1})}
\makeatother

\begin{document}
\begin{frontmatter}

\title{Estimating population size using the network scale~up method\thanksref{T1}}
\runtitle{Estimating population size}

\begin{aug}
\author[A]{\fnms{Rachael}~\snm{Maltiel}\thanksref{M1}\ead[label=e1]{rmaltiel@uw.edu}},
\author[B]{\fnms{Adrian E.}~\snm{Raftery}\corref{}\thanksref{M2}\ead[label=e2]{raftery@uw.edu}},
\author[B]{\fnms{Tyler~H.}~\snm{McCormick}\thanksref{M2}\ead[label=e3]{tylermc@uw.edu}}
\and
\author[B]{\fnms{Aaron J.}~\snm{Baraff}\thanksref{M2}\ead[label=e4]{ajbaraff@uw.edu}}
\runauthor{Maltiel, Raftery, McCormick and Baraff}
\affiliation{Expedia\thanksmark{M1} and University of Washington\thanksmark{M2}}
\address[A]{R. Maltiel\\
Marketing Analysis\\
Expedia, Inc.\\
333 108th Avenue NE\\
Bellevue, Washington 98004\\
USA\\
\printead{e1}}
\address[B]{A. E. Raftery\\
T. H. McCormick\\
A. J. Baraff\\
Department of Statistics\\
University of Washington\\
Box 354322\\
Seattle, Washington 98195-4322\\
USA\\
\printead{e2}\\
\phantom{E-mail: }\printead*{e3}\\
\phantom{E-mail: }\printead*{e4}}
\end{aug}
\thankstext{T1}{Supported in part by the Eunice Kennedy Shriver
National Institute of Child Health and Development (NICHD)
Grants R01 HD054511 and R01 HD070936,
U.S. Army Research Office Grant 62389-CS-YIP and a Science Foundation
Ireland E. T. S.~Walton visitor award, grant reference 11/W.1/I2079.}

\received{\smonth{8} \syear{2013}}
\revised{\smonth{12} \syear{2014}}

\begin{abstract}
We develop methods for estimating the size of hard-to-reach populations
from data
collected using network-based questions on standard surveys. Such data
arise by asking respondents how many people they know in a specific
group (e.g., people named Michael, intravenous drug users).
The Network Scale up Method (NSUM) is a tool for producing population size
estimates using these indirect measures of respondents' networks.
Killworth et~al. [\textit{Soc. Netw.} \textbf{20} (1998a) 23--50,
\textit{Evaluation Review} \textbf{22} (1998b) 289--308]
proposed maximum
likelihood estimators of population size for a fixed effects model in which
respondents' degrees or personal network sizes are treated as fixed.
We extend this by treating personal network sizes as random effects,
yielding principled statements of uncertainty.
This allows us to generalize the model to account for variation
in people's propensity to know people in particular subgroups
(barrier effects), such as their tendency to know people like themselves,
as well as their lack of awareness of or reluctance to acknowledge
their contacts' group memberships (transmission bias).
NSUM estimates also suffer from recall bias,
in which respondents tend to underestimate the number of members of
larger groups that they know, and conversely for smaller groups.
We propose a data-driven adjustment method to deal with this.
Our methods perform well in simulation studies, generating
improved estimates and calibrated uncertainty intervals, as well as in
back estimates of real sample data.
We apply them to data from a study
of HIV/AIDS prevalence in Curitiba, Brazil.
Our results show that when transmission bias is present, external
information about its likely extent can greatly improve the estimates.
The methods are implemented in the NSUM R package.
\end{abstract}

\end{frontmatter}

\section{Introduction} \label{introduction}

The problem of estimating the size of hard-to-reach subpopulations
arises in many contexts.
In countries with concentrated HIV/\break AIDS epidemics, the sizes of
key affected populations are important for estimating and projecting
the epidemic.
Concentrated AIDS epidemics are defined as epidemics where AIDS is
largely concentrated within particular at-risk groups, such as
intravenous drug users (IDU), female sex workers (FSW) and men who have
sex with men (MSM).
Estimates of the sizes of these groups are also needed to appropriately
distribute resources and prevention programs to contain the AIDS epidemic.

The Network Scale Up Method (NSUM) has been proposed as a way to
estimate the size of hard-to-reach subpopulations. The NSUM was first
proposed by \citeauthor{bernard1989estimating}
(\citeyear{bernard1989estimating,bernard1991estimating})
following the 1985 Mexico City earthquake in an attempt to use
respondents' knowledge about their social contacts to estimate the
number of people that died in the earthquake. Bernard and colleagues
realized that the information an individual possesses about others in
his or her social network could be used to estimate populations that
are currently difficult to size.

Respondents are asked questions of the type ``How many $X$ do you
know?,'' where $X$ ranges over different subpopulations of both known
and unknown size. Known subpopulations could include people named
Michael, diabetics and women who gave birth to a baby, while unknown
subpopulations are typically the groups of interest, such as female sex
workers. To standardize what it means to know someone, the
\citet{mccarty2001comparing} survey defines it as follows: ``For the purposes
of this study, the definition of knowing someone is that you know them
and they know you by sight or by name, that you could contact them,
that they live within the United States and that there has been some
contact (either in person, by telephone or mail) in the past 2 years.''
The survey can be applied to anyone in the overall population of
interest. Respondents do not have to admit to belonging to any
particular group, unlike in
most other survey methods.
``How many $X$ do you know?'' questions can easily be integrated into
almost any survey,
allowing the method to be implemented with limited cost.

Previous statistical work in this area refers to ``How many $X$ do you
know?'' data as aggregated relational data. These questions are widely
used on surveys such as the General Social Survey to measure
connectivity patterns between individuals. Statistical work in this
area includes \citet{zheng2006many} who used
aggregated relational data to estimate social structure through
overdispersion, \citet{mccormick2010many} who developed methods for
estimating individuals' personal network size and rates of mixing
between groups in the population, and \citet{McCormick2012tq} who
estimated the demographic composition of hard-to-reach populations.
While we focus here on estimating the sizes of population groups, the
previous work focused primarily on estimating features of the
population social network and the dynamics of interactions between
population groups.

In its simplest form, the NSUM is based on the idea that for all
individuals, the probability of knowing someone in a given
subpopulation is the size of that subpopulation divided by the overall
population size. For example, if a respondent knows 100 people total
and knows 2 intravenous drug users, then it is inferred that 2\% of the
total population are intravenous drug users. This assumption
corresponds to a binomial model for the number of people in a given
subpopulation that the respondent knows. However, the total number of
people known by a respondent, also called his or her degree or personal
network size, also needs to be estimated. A person's degree is
estimated by asking the respondents about the number of contacts he or
she has in several subpopulations of known size, such as twins, people
named Nicole or women over 70, using the same assumption that an
individual should know roughly their degree times the proportion of
people in a given subpopulation. The size of the unknown subpopulation
is then estimated using responses to questions about the number of
people known in the unknown subpopulation combined with the degree
estimate, leading to the scale-up estimator
[\citeauthor{killworth1998social} (\citeyear{killworth1998social,killworth1998estimation})].
The estimator can be improved by increasing the number of respondents
and the number of known subpopulations asked about.

The scale-up estimator suffers from several kinds of bias
[\citeauthor{killworth2003two} (\citeyear{killworth2003two,PKetal06}), \citet{mccormick2010many}].
It does not take account of the different propensities of people
to know people in different groups,
such as people's tendency to know people like themselves;
these are called barrier effects.
Transmission bias arises when a respondent does not count his or her contact
as being in the group of interest, for example, because the respondent
does not know that the contact belongs to the group.
This bias may be particularly large when
a group is stigmatized, as is the case of most of the key affected
populations in which we are interested.
Recall bias refers to the tendency for people to underestimate the
number of
people they know in larger groups because they forget some of these
contacts, and to overestimate the number of people they know in small
or unusual groups.

\citet{mccormick2010many} proposed strategies for improving degree
estimation. Efficiently estimating respondent degree was the focus of
that work, however, and so it did not directly address estimating
population size. Further, the \citet{mccormick2010many} method
requires additional information about the demographic composition of
populations with known size. This information is not always available
when estimating population group size.
Similarly, \citet{mccormick2007adjusting} proposed a calibration curve to adjust for
recall bias that was later incorporated
into \citet{mccormick2010many}. We use a similar approach to address recall
issues, but adjust our approach to ensure compatibility with our model
for size estimation.

Some attempts have been made to correct for transmission bias in size
estimates.
These consist of estimating the probability that a respondent counts
a contact that belongs to the group of interest as being a member of
the group,
and then dividing the NSUM size estimate by the estimated probability.
\citet{ezoe2012population}
surveyed men who have sex with men, the population of interest, to find
out how many people in the MSM's networks knew about their group status.
Salganik et~al.'s (\citeyear{salganik2011game}) implementation of NSUM estimates in
Curitiba, Brazil included a game of contacts method where the
researchers surveyed heavy drug users to estimate the proportion of
their network that are aware of their drug use status. The game of
contacts method involves asking heavy drug users about the number of
people they know with certain names and then asking if those contacts
are aware of the respondent's drug use status as well as the contacts'
own drug use status. This allows for an estimate of the proportion of
drug users that NSUM survey respondents would be aware of within their
own social network. The success of these methods remains to be determined.

\citeauthor{zheng2006many}'s (\citeyear{zheng2006many}) model involved a parameter denoted by~$b_k$, defined as the prevalence parameter or the proportion of total
links that involve group $k$, and they provided a way of estimating it.
It is tempting to interpret this as the proportion of the population
in group $k$, and hence as providing a population size estimate for
group $k$.
However, this is incorrect, particularly for populations for
which transmission bias is a major concern, such as the hard-to-reach
populations that are our main focus. If \citeauthor{zheng2006many}'s (\citeyear{zheng2006many}) prevalence
parameter $b_k$ was used to estimate the size of hard-to-reach populations,
it would tend to give substantially biased estimates.

In this paper, we develop a Bayesian framework for population group
size estimation using the NSUM.
We first build a random degree model with a random effect for degree
which incorporates variability and uncertainty across individuals'
network sizes.
We then build on this basic model to adjust for barrier and transmission
effects, both separately and combined, resulting in four models altogether.
The overall goal is to provide size estimates with reduced bias and error,
as well as to assess the uncertainty of the estimates.
The methods developed are implemented in the freely available
NSUM R package.

In Section~\ref{models} we introduce the four models: the random
degree model, the barrier effect model, the transmission bias model,
and the combined barrier effect and transmission bias model. We also
propose a method for adjusting for recall bias.
In Section~\ref{results} we show results from several simulation
studies, confirming the need to account for biases and the success of
our methods
in correcting for them.
We also show that adjusting for barrier effects using our methods yields
better size estimates than the
\citeauthor{killworth1998social}
(\citeyear{killworth1998social,killworth1998estimation})
estimates for the known populations
in the data set used by \citet{mccarty2001comparing}.
We will also show the estimates produced by our model on the
Curitiba, Brazil data of \citeauthor{salganik2011assessing} (\citeyear{salganik2011assessing,salganik2011game}).
Last, in Section~\ref{discussion} we will discuss additional research
needed to make NSUM estimation a viable, accurate method to estimate
the size of hard-to-reach populations.

\section{Models} \label{models}
Previous size estimates based on ``How many $X$'s do you know?'' data
have been computed using the
network scale-up estimator. Let $y_{ik}$ be the number of people known by
individual $i$, $i=1,\ldots,n$, in group $k$, $k=1,\ldots,K$,
with groups $1,\ldots,K-1$ being of known size and
group $K$ of unknown size. (Note that there can be more than one group
of unknown size, but we are using one to simplify the exposition.)
Let $d_i$ be the number of people that respondent $i$ knows,
also called his or her degree or personal network size.
Also, let $N_k$ be the size of group $k$, and let $N$ be the total population,
which is taken to be known.

The scale-up estimates are based on the assumption that
$y_{ik} \sim\break \operatorname{Binom}(d_i, \frac{N_K}{N})$,
or that the number of people known by individual $i$
in group $k$ follows a binomial distribution. We refer to this as the
scale-up model. From this model,
\citeauthor{killworth1998social} (\citeyear{killworth1998social,killworth1998estimation})
derived the maximum likelihood estimator of $d_i$ as
%
\begin{equation}
\label{killworth.di}
\hat{d}_i = N \frac{ \sum_{k=1}^{K-1} y_{ik}}{\sum_{k=1}^{K-1} N_k}.
\end{equation}
Conditional on estimates $\hat{d}_i$ of $d_i$, the maximum likelihood
estimator of $N_K$, the size of the unknown population, is then
%
\begin{equation}
\label{killworth.NK}
\hat{N}_K = N \frac{ \sum_{i=1}^{n} y_{iK}}{\sum_{i=1}^{n}
\hat{d}_i}.
\end{equation}
Equations \eqref{killworth.di} and \eqref{killworth.NK} are commonly
referred to as the scale-up estimates.

\begin{figure}

\includegraphics{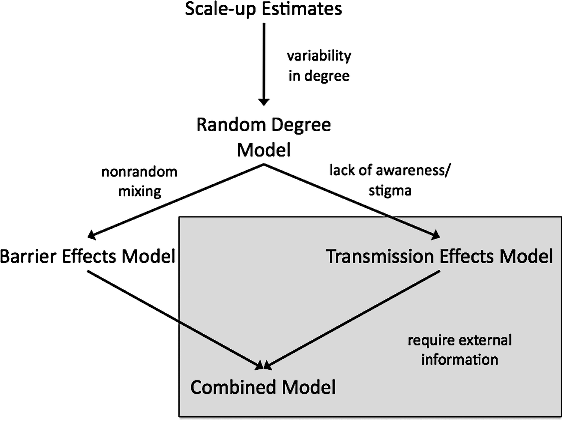}

\caption{Our four models build on the basic
\citeauthor{killworth1998social}
\textup{(\citeyear{killworth1998social,killworth1998estimation})}
scale-up model, accounting for
nonrandom mixing or barrier effects, and transmission bias.}
\label{model.diagram}
\end{figure}

Our proposed models build on the scale-up model.
We first model degree as a random effect, leading to regularized
estimates of degree. We refer to this as our random degree model.
We then extend the random degree model to take account of the fact that
respondents
have different propensities to know members of different groups.
For example, people are generally more likely to know people that are similar
to them in terms of age, sex, education, race and other characteristics,
than to know people who are not.
We account for this nonrandom mixing of individuals with an additional random
effect, to yield what we call the barrier effects model.
We also separately extend the random degree model to account for lack
of awareness of or reluctance to acknowledge contacts' group memberships,
to yield what we call the transmission bias model.
We find that the quality of estimates from this model can be greatly
improved by
external information on information transmission.
Last, our combined model accounts for both barrier effects and
transmission bias.
The models build on each other, as described in Figure~\ref{model.diagram}.

\subsection{Random degree model} \label{secrandom.degree.model}

Our first extension of the
\citeauthor{killworth1998social} (\citeyear{killworth1998social,killworth1998estimation})
scale-up model is to introduce
a random effect for degree, to regularize estimates of degree.
If an individual responded that he or she knew a large number of people
in a given subpopulation, this would drive up the estimate of the
individual's degree $d_i$.
To reduce the sensitivity of estimates to extreme values of $d_i$,
we incorporate degree estimation into our hierarchical modeling
framework and
achieve regularization through partial pooling.

We call the resulting model our random degree model. It assumes that
\begin{eqnarray*}
y_{ik} &\sim & \operatorname{Binom} \biggl( d_i ,
\frac{N_k}{N} \biggr),
\\
d_i &\sim & \operatorname{Log \ Normal} \bigl(\mu, \sigma^2\bigr).
\end{eqnarray*}
We choose a log normal distribution for $d_i$ based on the observed
distribution of scale-up estimates of degree $\hat{d}_i$. We found
the log normal distribution to have the best fit to estimates of
$\hat{d}_i$ across multiple data sets, including data from the
United States, Ukraine, Moldova, Kazakhstan and Brazil [\citet
{mccarty2001comparing,paniotto2009estimating,salganik2011assessing}].
We estimate the parameters of the random degree model in a Bayesian manner,
using the prior distributions
\begin{eqnarray*}
\pi(N_K) &\propto & \frac{1}{N_K} 1_{N_K \leq N},
\\
\mu & \sim & \mathit{U}(3,8),
\\
\sigma&\sim & \mathit{U}\biggl(\frac{1}{4}, 2\biggr). 
\end{eqnarray*}
Our prior for $N_K$ has been used previously for Bayesian estimation of
population size with little prior information
[\citet{Jeffreys1961,Raftery1988}].
The priors for $\mu$ and $\sigma$ were arrived at from the values we
saw in fitting the scale-up $\hat{d}_i$ estimates to several data
sets across multiple regions. Our prior for $\mu$ allows for mean
degrees within a data set ranging from 20 to 3000, which is consistent
with previous research on social networks and the NSUM [\citet{mccarty2001comparing,mccormick2010many}]. Our prior on $\sigma$
allows for 95\% of degrees to fall in the multiplicative range 1.6
times to 55 times in either direction from the mean, which seemed to
more than fully cover the range of results from scale-up estimates
across multiple data sets.

\subsection{Barrier effects model} \label{secbarrier.model}

Nonrandom mixing, or barrier effects, occur because respondents have different
tendencies to know people in different groups, depending on their
own characteristics.
For example, we might expect a 65-year-old male respondent to know more
people named Walter than a 20-year-old female respondent, because
Walter was a more common name 65 years ago.
This leads to overdispersion in the distribution of the number of people
known in a given population relative to what one would expect if the
binomial assumption held.

We can model overdispersion in the binomial probabilities as follows.
In the \citeauthor{killworth1998social}
(\citeyear{killworth1998social,killworth1998estimation})
scale-up and random degree models, the probability that respondent $i$
knows someone in group $k$ is assumed to be constant across respondents,
and equal to $N_k/N$. To model overdispersion, we instead allow
this probability, now denoted by $q_{ik}$, to vary randomly across respondents,
following a Beta distribution.
The model then becomes
\begin{eqnarray*}
y_{ik}  &\sim & \operatorname{Binom} (d_i , q_{ik}),
\\
d_i &\sim & \operatorname{Log \ Normal} \bigl(\mu,
\sigma^2\bigr),
\\
 q_{ik} & \sim &  \operatorname{Beta} (m_k, \rho_k).
\end{eqnarray*}
Here we use the nonstandard parameterization of the Beta distribution
according to which $X \sim\operatorname{Beta} (m, \rho)$ if it has the
probability density function $f_X(x) \propto x^{\alpha-1} (1-x)^{\beta-1}$,
where
$m = \frac{\alpha}{\alpha+\beta}$ and $\rho= \frac{1}{1 + \alpha
+ \beta}$
[\citet{diggle2002analysis}, Chapter~9,\vspace*{1pt}  \citet{skellam1948probability,mielke1975convenient}].
Then $m_k$ is the prior mean of $q_{ik}$, and $\rho_k$ determines
its dispersion. We set $E[q_{ik}] = m_k = \frac{N_k}{N}$.
We use the prior distributions
\begin{eqnarray*}
\pi(m_K) & \propto & \frac{1}{m_K},
\\
\rho_k &\sim & \mathit{U}(0,1),
\end{eqnarray*}
with the priors for $\mu$ and $\sigma$ remaining the same as in the
random degree model.

\subsection{Transmission bias model} \label{sectransmission.model}

Transmission bias occurs when a respondent is unaware of or reluctant to
acknowledge the group membership status of his or her contacts. For
example, if a respondent is not aware that a contact is an intravenous
drug user, he or she would not count that contact when responding to a
question about the number of intravenous drug users known. We can think
of the transmission bias, denoted
by $\tau_k$, as the proportion of respondents' contacts in group $k$
that the respondents report. For example, if 50\% of intravenous drug
users disclose their status to their contacts and if respondents report
all the IDUs that they know,
then $\tau_K = 0.5$ for the subpopulation $K$ of IDUs.
Thus, we can add $\tau_k$ to our model as a multiplier of the binomial
proportion, since a respondent would mention knowing only a proportion
$\tau_k$ of their true contacts in group $k$ on average.
This yields the transmission bias model
\begin{eqnarray*}
y_{ik}  &\sim & \operatorname{Binom} \biggl( d_i ,
\tau_k \frac{N_k}{N} \biggr),
\\
d_i  &\sim & \operatorname{Log \ Normal} \bigl(\mu,
\sigma^2\bigr).
\end{eqnarray*}

We specify the additional prior
\[
\tau_K \sim\operatorname{Beta}(\eta_K, \nu_K),
\]
with the priors for $N_K, \mu$ and $\sigma$ remaining the same as in
the random degree model. For the transmission bias, we assume $\tau_k$
to be 1 for the known populations $k=1,\ldots,K-1$, and to be less
than or equal to
one for the groups of unknown size,
in line with the definition of transmission bias.
This means that we are assuming that respondents are aware of and prepared
to acknowledge contacts' group membership status for the known groups.
This assumption is reasonable, as the known populations are typically
less stigmatized, making it less likely for respondents to be unaware
of or reluctant to
acknowledge their contacts' membership statuses.
Our simulation results indicated the desirability of using external information
about $\tau_K$ in the form of an informative prior,
which will be discussed further in Section~\ref{secsim.results}.

\subsection{Combined model} \label{seccombined.model}

Previous research indicates both barrier effects and transmission bias
to be present in these data [\citet{mccarty2001comparing,kadushin2006scale,mccormick2010many,salganik2011assessing}]. For a
model to produce unbiased estimates, we need to adjust for both sources
of bias. Thus, we can combine our barrier and transmission models to
get a combined model that accounts for both barrier effects and
transmission bias. Our model is thus
\begin{eqnarray*}
y_{ik} &\sim & \operatorname{Binom} (d_i , \tau_k
q_{ik}),
\\
%
d_i  &\sim & \operatorname{Log \  Normal} \bigl(\mu,
\sigma^2\bigr),
\\
q_{ik}  &\sim & \operatorname{Beta} (m_k, \rho_k),
\end{eqnarray*}
with priors the same as in the previous models.

\subsection{Recall bias adjustment} \label{recall.adjustment}

Since respondents are asked to say quickly how many people they know in
certain groups, it is common for them to forget contacts in large
groups or to overcount contacts in small groups. For example, a
respondent might know 15 or 20 people in
a large group and might forget to mention a few while quickly answering
a survey.
In addition, small subpopulations can be memorable, such as people who
died in a car accident. Respondents might count someone in a small
subpopulation as someone they know even if the contact does not
actually fall under the definition of ``know'' in NSUM surveys.

Previous research has suggested methods to adjust for recall bias based
on the relationship between respondents' recalled ties and the sizes of
known groups of interest [\citet{killworth2003two,zheng2006many,mccormick2007adjusting,mccormick2010many}].
Our exploratory work suggests a linear relationship between the two on
the log scale.
This leads to the following model to incorporate recall bias as well as
barrier effects and transmission bias:
\begin{eqnarray*}
y_{ik} & \sim & \operatorname{Binom} \bigl(d_i , e^{r_k}
\tau_k q_{ik}\bigr),
\\
%
r_k  &\sim &  N\bigl(a + b \log N_k,
\sigma_r^2\bigr),
\\
d_i  &\sim & \operatorname{Log \ Normal} \bigl(\mu, \sigma^2\bigr),
\\
q_{ik}  &\sim & \operatorname{Beta} (m_k, \rho_k).
\end{eqnarray*}
The additional parameters $a, b$ and $\sigma_r$ have uniform flat priors,
namely, $a \sim U(0,15)$, $b \sim U(0,1)$ and $\sigma_r \sim U(0, 1)$.
The quantity $N_k$ would be calculated just as in the barrier and
combined models, where $N_k = N \cdot m_k$.

However, this model involves a large number of parameters and is
quite computationally demanding. For models estimating one unknown
subpopulation, the random degree model has $n+3$ parameters, the
barrier model has $n+K+2$ parameters, and the transmission model has
$n+4$ parameters. This full model has $n+2 K + n\cdot K + 7$ parameters---a large increase from the simpler models. This increase in
parameters, coupled with the limited information about recall bias
present in the data, makes inference for this model difficult and, in
our judgment, not a worthwhile investment. Instead, we approximate a
recall adjustment through a postprocessing method. This method is
computationally very efficient and makes effective use of information
available through populations with known size. This method is also
easier to implement and, thus, improves the likelihood that the method
will be used in practice. The barrier and transmission combined model
similarly has $n + K + n \cdot K + 4$ parameters, but the relationship
between barrier effects and transmission bias makes a similar
postprocessing approach difficult in this case.

We outline our recall adjusted modeling strategy below. We found that
this strategy performed well in practice in our data experiments. We
first estimate a linear relationship (on the log scale) between the
estimates and the true subpopulation sizes using back estimates.
For a data set with $K-1$ known subpopulations, back estimates estimate
the $k$th subpopulation, $k=1, \ldots, K-1$, treating it as
unknown, and treating all other $K-2$ known subpopulations as known to
produce the estimate. This can be done for all $K-1$ known
subpopulations and then compared to the true, known sizes of those
subpopulations for estimation method evaluation.
To account for the variability in our estimate of $\hat{N}_k$ as well,
we approximate the relationship using
the errors-in-variables model
%
\begin{equation}
\label{recall.adjust.eqn}
\log(\hat{N}_k) = a + b \log(N_k) +
\delta_k + \varepsilon_k,
\end{equation}
where $\hat{N}_k$ is the posterior mean and $s_k$ the posterior
standard deviation of the size of the $k$th subpopulation,
computed without knowledge
of the true $N_k$, $\delta_k \sim N(0, s_k^2)$, and
$\varepsilon_k \sim N(0,\sigma^2_{\varepsilon})$.
The model (\ref{recall.adjust.eqn}) is estimated by maximum likelihood
[\citet{ripley1987regression}].

We then adjust for recall bias as follows.
Let $Y_K^{[t]}$ denote the $t$th value simulated from the posterior
distribution of $\log(N_K)$, where $t$ indexes MCMC iterations.
We then replace each $Y_k^{[t]}$ with a randomly drawn value
\[
\frac{Y_K^{[t]} - a}{b} + Z,
\]
where $Z \sim N(0, \sigma_{\varepsilon}^2 / b^2)$ to adjust for
recall bias,
based on the relationship shown in equation \eqref{recall.adjust.eqn}.
In our analyses, we have generally found $a$ to be around 6.7,
$b$ to be around 0.5, and $\sigma_{\epsilon}$ to be around 0.35. Our
strategy differs from that of~\citet{mccormick2007adjusting}
and~\citet{mccormick2010many} because we apply our adjustment after a
complete run of our sampler. The correction for recall cannot,
therefore, influence the path of the sampler as in~\citet{mccormick2007adjusting} and~\citet{mccormick2010many}. The strategy
is instead more similar to that employed by~\citet{zheng2006many}, who
adjusted a normalization constant (necessary to preserve
identifiability) after sampling to adjust for recall issues. Our
proposed method propagates uncertainty from responses to size
estimates, however, which is not a feature of the~\citet{zheng2006many} approach.

\section{Results} \label{results}
We estimated all the models using Markov chain Monte Carlo (MCMC).
For all models, $\mu$ and $\sigma$ were sampled from using
closed-form Gibbs
steps while we used random walk Metropolis steps with normal proposals
for all the other
parameters. Derivations of all Gibbs and Metropolis steps are included
in the \hyperref[app]{Appendix}. When possible, we used scale-up estimates as starting
points for the parameters.

The MCMC algorithms were implemented using the methodology described in
\citet{raftery1996implementing}, using an initial chain to estimate
the conditional
posterior standard deviation of each parameter given the other parameters,
and then using 2.3 times this value as the standard deviation in the
normal proposal. We used the Raftery--Lewis diagnostic to determine the
number of iterations needed for the MCMC. In general, our chains
behaved well, converging in less than 30{,}000 iterations. Our combined
model, though, required over 150,000 iterations. We also checked the
Gelman--Rubin diagnostic on all models on the Curitiba data set,
discussed below [\citet{gelman1992inference}]. For $N_K$, our population
size of interest, the Gelman--Rubin diagnostic was close to 1 in all
models. For the other parameters, the Gelman--Rubin diagnostic was under
1.015 in the random degree, barrier and transmission models and under
1.1 for 99.5\% of the  10{,}416 parameters in the combined model.

One difficulty in verifying NSUM estimation results is that we do not
know the true size of hard-to-reach subpopulations. Thus, we first ran
several simulations to verify the need for and improvement from our
models that adjust for biases when present. We tested our models on
data containing no bias, barrier effects and transmission bias for
three types of simulations and we report the results in Section~\ref{secsim.results}. Second, we computed back estimates on the data from
\citet{mccarty2001comparing}, or estimates of known subpopulations to
be compared to the true size, to assess the efficacy of our models,
detailed in Section~\ref{secmccarty.back}. Last, in Section~\ref{seccuritiba} we give
results from estimating all our models on data from the
Curitiba study
[\citeauthor{salganik2011assessing} (\citeyear{salganik2011assessing,salganik2011game})].

\subsection{Simulation studies} \label{secsim.results}

For our simulations, we created data sets containing various effects
and biases: no effects or biases, barrier effects, transmission bias,
and both barrier effects and transmission bias. In the simulations with
no effects or biases, the data followed the assumptions of our random
degree model: the respondents' degrees followed a log normal
distribution, while the number of people known in each group followed a
binomial distribution based on the respondent's degree and the
proportion of the total population in a given group. In the barrier
effects simulations, we added a beta random effect to the binomial
proportion. For the data with transmission bias, we instead added a
multiplier $\tau_K$ to the binomial proportion. To simulate data with
both biases, we added the beta random effect and the multiplier $\tau
_K$ to the binomial proportion.

The simulations with no effects or biases and with only barrier effects
were based on data from \citet{mccarty2001comparing}, while the
simulations including a transmission bias were based on data from
\citet{salganik2011assessing}. While the \citet{mccarty2001comparing}
data is a well understood and commonly used data set, we had more
detailed information on transmission bias for the prior in the
\citet{salganik2011assessing} Curitiba data set, making it a better choice on
which to base a transmission bias simulation. For all simulations, we
used a sample size of 500 and simulated 100 data sets. We estimated the
size of one unknown population; for the simulations based on
\citet{mccarty2001comparing}, the unknown population had size 500{,}000 [based
on scale-up estimates of the unknown groups in the \citet{mccarty2001comparing} data set], while for the simulations based on
\citet{salganik2011assessing}, the unknown population had size 65{,}000
(based on the scale-up estimates of heavy drug users in Curitiba). When
barrier effects were present in the data, we used values for the
barrier effect parameters estimated in the \citet{mccarty2001comparing} data set by the barrier effect model. For
transmission bias, we used $\tau_K = 0.54$ based on the estimate of
transmission bias from \citet{salganik2011game} using the game of
contacts method. We also obtained our transmission bias prior of
$\operatorname{Beta}(0.542, 0.011)$ by fitting a beta distribution to the bootstrapped
estimates of the transmission bias $\tau_K$.

\begin{figure}

\includegraphics{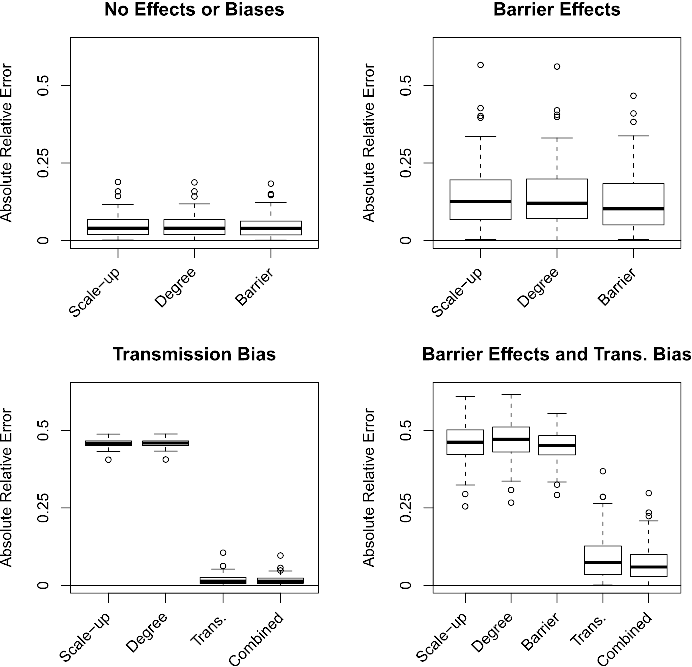}

\caption{Simulation study: absolute relative errors (ARE) of posterior means
of $N_K$ relative to the true size of $N_K$.
Each panel corresponds to a different simulation setup.
The four boxplots in each panel correspond to different estimates:
scale-up estimates, random degree model estimates, and estimates
from either the barrier effects model, the transmission bias model, or
the combined model.
Each boxplot shows the distribution of the AREs across 100 simulated
data sets.}
\label{mae}
\end{figure}

\citet{salganik2011game} used both a transmission bias parameter, to
measure respondents' awareness of contacts' status, and a differential
network size parameter, to measure differences in the size of networks
of people in the population of interest versus people in the general
population. We have combined these two parameters for our transmission
bias parameter, as they are not identifiable in our models without an
additional sample of individuals from the population of interest to
perform the game of contacts.

Across our simulations, we measured the mean absolute relative error
(MARE) to see how much error occurred in estimates when using different
models based on different assumptions. Figure~\ref{mae} depicts the
absolute errors scaled by the true size of the unknown population, with
the point estimate being the mean of the posterior of~$N_K$, while the
numbers are reported in Table~\ref{tabmae} as well. We see that when
there are no barrier effects or biases in the data, the scale-up
estimates and random degree model produce estimates with little error.
The barrier effects model is also able to estimate population size with
minimal error,
even though the barrier effects that the model includes are not present
in the data.
When barrier effects are present in the data, the barrier effects model
produces a MARE that is 12\% lower than the scale-up estimates or
the random degree model.

We see the largest difference in estimates when transmission bias is
present in the data. When transmission bias is not accounted for in the
model estimates, the MARE is large, while the transmission and combined
models result in estimates with minimal error. When both barrier
effects and transmission bias are present, the combined model produces
a MARE that is 21\% lower than the estimates that account for
transmission bias alone. Interestingly, the combined model results in
slightly lower MARE even when no barrier effects are present in the data.

\begin{sidewaystable}
\tabcolsep=0pt
\tablewidth=\textwidth
\caption{Mean absolute relative error, relative to the true
subpopulation sizes, and coverage over the 100 simulations across data
set designs and estimation models: scale-up model, random degree
(Degree) model, barrier effects model, transmission bias (Trans.)
model, and combined model}\label{tabmae}
\begin{tabular*}{\tablewidth}{@{\extracolsep{\fill}}lccccccccccccccc@{}}
\hline
\textbf{Data} & \multicolumn{3}{c}{\textbf{No effects or biases}} & \multicolumn
{3}{c}{\textbf{Barrier effects}} & \multicolumn{4}{c}{\textbf{Transmission bias}} &
\multicolumn{5}{c@{}}{\textbf{Barrier effects and trans. bias}} \\[-4pt]
& \multicolumn{3}{l}{\hrulefill} & \multicolumn{3}{l}{\hrulefill}
& \multicolumn{4}{l}{\hrulefill} & \multicolumn{5}{l@{}}{\hrulefill}\\
\textbf{Model} & \textbf{Scale-up} & \textbf{Degree} & \textbf{Barrier} &
 \textbf{Scale-up} & \textbf{Degree} & \textbf{Barrier} &
\textbf{Scale-up} & \textbf{Degree} & \textbf{Trans.} & \textbf{Combined} &
\textbf{Scale-up} & \textbf{Degree} & \textbf{Barrier} &
\textbf{Trans.} & \textbf{Combined} \\
\hline
MARE & 0.046 & 0.046 & 0.046 & 0.145 & 0.145 & 0.128 & 0.459 & 0.459 &
0.018 & 0.017 & 0.462 & 0.471 & 0.447 & 0.091 & 0.072 \\
MARE SE & 0.003 & 0.003 & 0.003 & 0.012 & 0.012 & 0.010 & 0.001 & 0.001
& 0.002 & 0.001 & 0.006 & 0.006 & 0.005 & 0.007 & 0.006 \\
80\% Coverage & -- & 84\% & 83\% & -- & 27\% & 87\% & -- & 0\% & 100\% &
85\% & -- & 0\% & 0\% & 74\% & 83\% \\
95\% Coverage & -- & 97\% & 97\% & -- & 48\% & 94\% & -- & 0\% & 100\% &
97\% & -- & 0\% & 0\% & 90\% & 91\% \\
\hline
\end{tabular*}
\end{sidewaystable}

Our credible interval coverage, shown in Table~\ref{tabmae}, also indicates
the value of using a model that correctly adjusts for bias in the data.
We see appropriate coverage for both the random degree and barrier
models when there is no bias in the data. When there are barrier
effects or transmission bias in the data, the random degree model
results in undercoverage, while the appropriate model yields accurate
interval coverage. In particular,
when transmission bias is present, the coverage of the random degree model
is close to zero.
While failing to account for barrier effects present in data results in
some error in estimates and undercoverage, the results are much more
extreme when failing to account for transmission bias. We believe
accurate assessment of transmission bias to be the highest priority in
improving NSUM size estimates.

Through our simulations, we were also able to see the importance of the
choice of priors for the transmission bias model. In addition to our
transmission bias simulation using the informative prior based on
\citeauthor{salganik2011game}'s (\citeyear{salganik2011game}) game of contacts results, we also ran a
simulation using $\operatorname{Uniform}(0,1)$ prior on $\tau_K$, which we will refer
to as an informative prior.
We found that for $\tau_K$, the posterior distribution
was very similar to the prior.
Table~\ref{tab2} gives the 95\% interval end points
and median for the $\tau_K$ prior as well as the average interval
endpoints and medians for the $\tau_K$ posterior for the simulations
with both informative and uninformative priors, where the posterior
values are averaged over the estimates from the 100 simulation
posteriors of $\tau_K$.

\begin{table}
\tablewidth=\textwidth
\caption{Comparison of prior and posterior 95\% credible interval
quantiles and medians for the uninformative and informative prior
transmission bias simulations, averaging over the posterior samples for
the 100 simulated data sets. We see that the posterior of $\tau_K$
aligns closely with the prior, showing the need for an informative
prior to produce accurate size estimates. In addition, we~see an
incorrect point estimate for prevalence using the uninformative prior,
and a wide range~of~uncertainty}\label{tab2}
\begin{tabular*}{\tablewidth}{@{\extracolsep{\fill}}lcccccc@{}}
\hline
& \multicolumn{3}{c}{\textbf{Transmission bias} $\bolds{\tau_K}$} & \multicolumn
{3}{c@{}}{\textbf{Prevalence}} \\[-4pt]
 & \multicolumn{3}{c}{\hrulefill} & \multicolumn{3}{c@{}}{\hrulefill}\\
 & \textbf{2.5\%} & \textbf{Median} & \textbf{97.5\%} & \textbf{2.5\%} & \textbf{Median} & \textbf{97.5\%}
 \\
\hline
Uninformative prior \\
\quad Prior & 0.025 & 0.500 & 0.975 & $5.5 \times10^{-5}$\% & 0.06\% &
68.8\% \\
\quad Posterior & 0.075 & 0.513 & 0.973 & 2.0\%\phantom{${}\times10^{-5}$} & 3.9\%\phantom{0} & 30.1\% \\[3pt]
Informative prior \\
\quad Prior & 0.438 & 0.542 & 0.644 & $5.5 \times10^{-5}$\% & 0.06\% & 68.8\%
\\
\quad Posterior & 0.438 & 0.542 & 0.644 & 3.0\%\phantom{${}\times10^{-5}$} & 3.6\%\phantom{0} & \phantom{0}4.5\% \\
\hline
\end{tabular*}
\end{table}

The close match between the prior and posterior of $\tau_K$ has major
implications for the posterior estimates of $N_K$ as well. Table~\ref{tab2} shows the 95\% credible interval points and
medians of $N_K$ averaged over the 100 simulations for both the
informative and uninformative prior as well.
The estimate of $N_K$ from the transmission bias model is roughly equal
to the estimate of $N_K$ from the random degree model divided by $\tau_K$.
Our estimates from the transmission bias model were very close to the
estimates in the random degree model divided by the prior expected
value $\tau_K$.
Thus, the error in the prior expectation of the transmission bias
will lead to a corresponding error in the estimate of $N_K$.
Our uninformative prior has an expected transmission bias, $\tau_K$,
of 50\%
(as compared to the true 54\%), and we do indeed see an overestimate of
the median prevalence in Table~\ref{tab2} when using
the uninformative prior: the true prevalence is 3.6\% as opposed to the
estimate of 3.9\%
with the noninformative prior.

If there is considerable uncertainty in the prior of $\tau_K$,
the posterior interval for $N_K$ will also be wide. The bottom two
panels of
Figure~\ref{mae} show the need to account for transmission bias
to produce an unbiased estimate, but Table~\ref{tab2}
indicates that an informative prior is needed to account for
transmission bias.
This indicates the need for methods to estimate transmission bias.

\subsection{McCarty back estimates} \label{secmccarty.back}

To further assess our methods,
we fit back estimates using the random degree and barrier effect models
for the
29 known subpopulations in the \citet{mccarty2001comparing} data set and
compared them to the known values.
In line with previous analyses, we assumed that there was no
transmission bias in these data, which seems reasonable given
these  are not stigmatized or hidden populations.
The \citet{mccarty2001comparing} data set was obtained through random
digit dialing within the United States. It contains responses from
1375 adults from two surveys: survey 1 with 801 responses conducted in
January 1998 and survey 2 with 574 responses conducted in January 1999.
The \citet{mccarty2001comparing} data set has been analyzed in
numerous articles, evaluating methods to estimate degrees in addition
to methods to estimate hard-to-reach populations [\citet{killworth2003two,zheng2006many,mccormick2010many}]. Since previous
research has indicated recall bias to be present in the McCarty data
set, we adjusted for recall bias as described in Section~\ref{recall.adjustment}.

\begin{figure}

\includegraphics{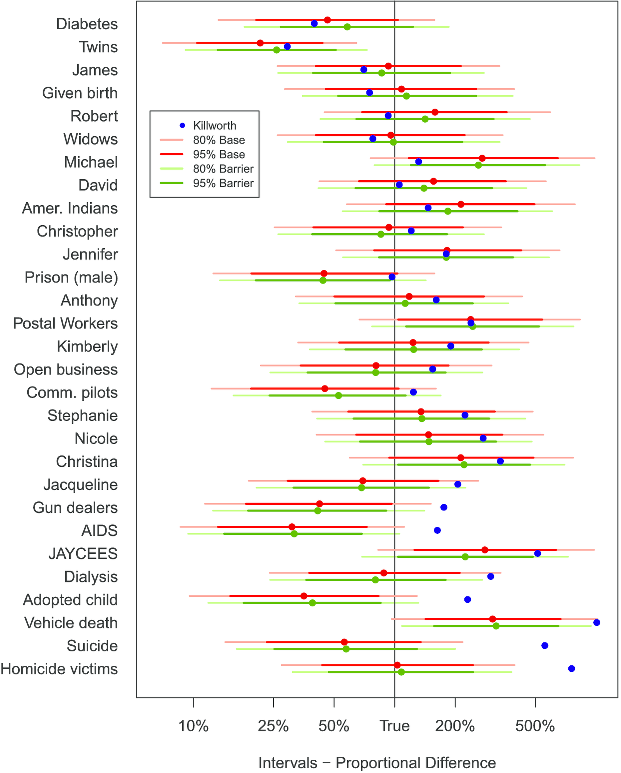}

\caption{Back estimates and 80\% and 95\% credible intervals for the
McCarty data sets using the random degree and barrier effect models and
scale-up estimates. The $x$-axis shows the estimates as proportions of
the true subpopulation sizes on the logarithmic scale, while the
$y$-axis shows the subpopulations in decreasing order of true size. The
black vertical line shows the goal where the estimates and true
subpopulation sizes are equal.}
\label{mccarty.intervals}
\end{figure}

Figure~\ref{mccarty.intervals} shows scale-up point estimates and
random degree model and barrier effects model
80\% and 95\% credible intervals of the posterior of the size estimates
of the \citet{mccarty2001comparing} data set shown as proportions of
the true subpopulation sizes. We see generally that our estimates are
close to the true subpopulation size and our credible intervals cover
the true subpopulation size.

\begin{figure}

\includegraphics{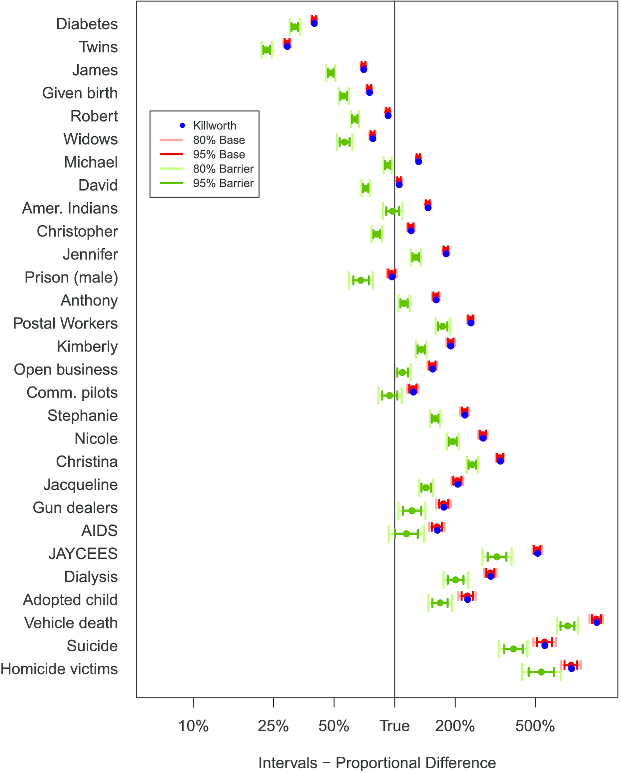}

\caption{Back estimates and 80\% and 95\% credible intervals for the
McCarty data sets using the random degree and barrier effect models and
scale-up estimates before recall bias adjustment. The $x$-axis shows
the estimates as proportions of the true subpopulation sizes on a log
scale, while the $y$-axis shows the subpopulations in decreasing order
of size. The black vertical line shows the goal where the estimates and
true subpopulation sizes are equal.}
\label{mccarty.intervals.no.recall}
\end{figure}

Figure~\ref{mccarty.intervals.no.recall} shows the same estimates and
credible intervals before adjusting for recall bias. We can see that
there is a clear association between recall bias and subpopulation size
and that the adjustment is important in correcting not only the
estimates but the credible intervals as well. It should also be noted
that unlike the method of~\citet{mccormick2010many}, our method
corrects for over-recall as well as under-recall, so good estimates can
be obtained for small subpopulations.

Table~\ref{mccarty.sum.stats} shows the mean absolute relative error
(MARE) and coverage of credible intervals for the estimation methods
over the 29 back estimates of the subpopulations in the \citet{mccarty2001comparing} data set. We see that the barrier model produces
estimates with the smallest average absolute relative error, as we
would hope given the barrier effects present in the McCarty data set.
We also see that both the random degree and barrier effects models
result in
accurate credible interval coverage.

\begin{table}
\tablewidth=240pt
\caption{Mean absolute relative error (MARE), standardized by dividing
all absolute errors by the true subpopulation sizes, and credible
interval coverage for scale-up estimates and random degree and barrier
model estimates over the 29 back estimates}\label{mccarty.sum.stats}
\begin{tabular*}{240pt}{@{\extracolsep{\fill}}lccc@{}}
\hline
& \multicolumn{3}{c@{}}{\textbf{Model estimates}} \\[-4pt]
& \multicolumn{3}{c@{}}{\hrulefill}\\
& \textbf{Scale-up} & \textbf{Degree} & \textbf{Barrier} \\
\hline
MARE & 1.49 & 1.48 & 0.93 \\
80\% Coverage & -- & 72\% & 66\% \\
95\% Coverage & -- & 97\% & 93\% \\
\hline
\end{tabular*}
\end{table}

\subsection{Curitiba results} \label{seccuritiba}

The Curitiba data set consists of 500 adult residents of Curitiba,
Brazil and was collected through a household-based random sample in
2010 by \citet{salganik2011assessing}.
One aim of this study was to estimate the sizes of hard-to-reach
populations relevant
to concentrated HIV/AIDS epidemics. In addition, a game of contacts survey
was conducted to estimate transmission bias for heavy drug users
[\citet{salganik2011game}]. From these game of contacts data,
we were able to obtain an informative prior for transmission bias,
allowing us to fit all of our models to the Curitiba data set and to
assess our models' performance on relevant data. As in our simulations,
we used a $\operatorname{Beta}(0.542, 0.011)$
prior for transmission bias based on the game of contacts estimate of
transmission bias.
We did not adjust for recall bias, as the study design did not produce
the information needed to do this.

\begin{figure}

\includegraphics{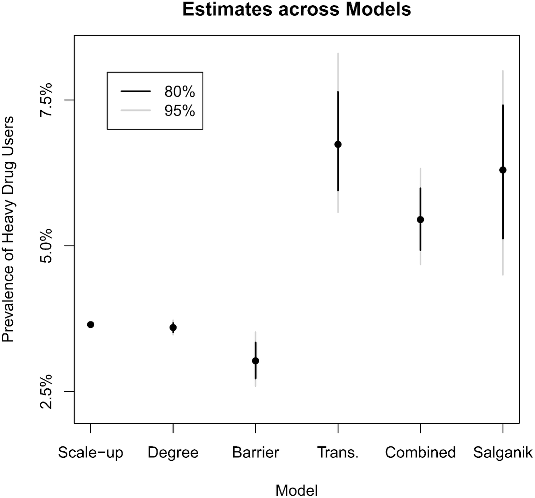}

\caption{Posterior estimates and credible intervals for the prevalence
of heavy drug users in Curitiba based on the random degree, barrier,
transmission and combined models, along with the \citeauthor{salganik2011assessing} \textup{(\citeyear{salganik2011assessing})} estimates after accounting for transmission bias.}
\label{curitiba.estimates}
\end{figure}

The estimates of prevalence of heavy drug users in Curitiba from our
models are shown in Figure~\ref{curitiba.estimates}. While there is
limited uncertainty in the estimates from the random degree model, the
estimates and their uncertainty are probably underestimated due to the
transmission bias in the data. The barrier model results in a smaller
estimate, while the transmission model results in a larger estimate of
heavy drug user prevalence. The uncertainty in the combined model seems
reasonable and is smaller than in the transmission model (and the
transmission prior) with a value between the separate barrier and
transmission model estimates. This compares to the estimates obtained
by \citet{salganik2011assessing} of 3.3\% with a 95\% confidence
interval from 2.7\% to 4.1\% without accounting for transmission bias,
and an estimate of 6.3\% with a 95\% confidence interval from 4.5\% to
8.0\% when accounting for transmission bias.


\section{Discussion} \label{discussion}

Indirectly observed social network data are one tool for estimating the
size of hard-to-reach populations. With knowledge of the true size of a
handful of subpopulations, data can be collected to then estimate the
size of hard-to-reach subpopulations that currently evade researchers.
These techniques can be used to provide accurate size estimates to
improve public health efforts related to AIDS in concentrated epidemics
as well as other subpopulations that are currently difficult to size.
NSUM surveys do not require large resources and can be carried out
by adding questions to other surveys already being conducted for other
purposes.

Currently the most used method for size estimation from such data is
the \citeauthor{killworth1998social} (\citeyear{killworth1998social,killworth1998estimation}) scale-up
estimate, but this does not provide estimates of uncertainty
and can suffer from barrier effects, transmission bias and recall bias.
In this paper we have proposed ways of overcoming these limitations.
First we proposed a Bayesian model, called the random degree model,
that regularizes estimation of degree
and yields estimates of uncertainty about population size.
Then we extended the model to incorporate barrier effects, transmission
bias and recall bias, and also proposed a more efficient postprocessing
method for accounting for recall bias.

We found that the barrier effects
model performs better than the scale-up estimates or the random degree model.
This makes sense because barrier effects, or nonrandom mixing,
are a pervasive feature of social networks. We also found that
adjusting for transmission bias is extremely important
when this bias is present. However, data typically do not contain
much information about transmission bias, and so it is important
to use or generate external information about transmission bias if possible.
Finally, we found that adjusting for recall bias can improve estimates
and the assessment of their uncertainty.

As seen in simulations in Section~\ref{secsim.results}, it is
important to adjust for bias in estimates through our proposed models
to minimize error in estimates and to produce appropriate coverage of
credible intervals. While nonrandom mixing can be accounted for using
our models that adjust for barrier effects without external
information, adjusting for transmission bias does require external
information. As seen in our simulations, since the posterior closely
aligns with the prior for the transmission bias effect, an informative,
accurate prior is needed to appropriately adjust estimates. While
researchers have started to find methods to estimate for transmission
bias, further work is needed in this area before NSUM can produce
estimates of hard-to-reach populations with an acceptable level of
error. The game of contacts of \citet{salganik2011game} is one way of
doing this.
The future utility of the NSUM will depend crucially on the
development and use of ways to estimate transmission bias.

We also observed that recall bias can only be effectively adjusted for
when the sizes of the known subpopulations encompass the size of the
unknown subpopulation. While the size of the unknown subpopulation is
unknown before estimation, researchers should aim to use external
sources to cover possible sizes of the group of interest.

\begin{appendix}\label{app}
\section*{Appendix: MCMC algorithms for model estimation}

This appendix contains derivations for the MCMC updates for the models
described in the main text. In Appendix~\ref{random.deg.deriv}, we have the derivations
for
the random degree model, detailed in Section~\ref{secrandom.degree.model}.
Appendix~\ref{A2} contains
the derivations for the barrier effects model, detailed in Section~\ref{secbarrier.model}.
The transmission bias model derivations are shown in Appendix~\ref{A3}, with
the model detailed in Section~\ref{sectransmission.model}. Last, Appendix~\ref{A4} contains
derivations for the combined model, which is detailed in Section~\ref{seccombined.model}.

\subsection{Random degree model} \label{random.deg.deriv}

The random degree model follows the binomial assumption of the
\citeauthor{killworth1998estimation}
(\citeyear{killworth1998social,killworth1998estimation}) model while adding a
random effect on degree to regularize degree estimates, as discussed in
Section~\ref{secrandom.degree.model}.
This yields the posterior distribution
\begin{eqnarray*}
&& \pi\bigl(\mu, \sigma^2, d_i, N_K |
y_{ik}, N_k, N\bigr)\\
&&\qquad \propto \prod
_{i=1}^{n} \Biggl[ \frac{1}{d_i \sigma\sqrt{2 \pi}} e^{-{(\log(d_i) - \mu)^2}/({2 \sigma^2})}
\prod_{k=1}^{K} \biggl( \pmatrix{d_i \cr y_{ik}} \biggl(\frac{N_k}{N}
\biggr)^{y_{ik}} \biggl(1-\frac
{N_k}{N} \biggr)^{d_i - y_{ik}} \biggr)
\Biggr]\\
&&\qquad\quad{}\times \frac{1}{N_K} \frac
{1}{5} \frac{1}{7/4}.
\end{eqnarray*}

First, $\sigma^2$ can be updated using a Gibbs sampler, as the
conditional posterior is a closed-form inverse gamma. Since $\sigma^2$
is inverse gamma, while
our prior is specified in terms of $\sigma$,
we need to include the Jacobian of the transformation, namely,
$\frac{1}{2} (\sigma^2)^{-1/2}$. 
The conditional posterior distribution of $\sigma^2$ is then
\begin{eqnarray*}
&& \pi\bigl( \sigma^2 | \mu, N_K, d_i,
y_{ik}, N_k, N\bigr) \\
&&\qquad\propto   \Biggl( \prod
_{i=1}^{n} \frac{1}{\sigma} e^{-{(\log(d_i) - \mu)^2}/({2
\sigma^2})} \Biggr)
\bigl(\sigma^2\bigr)^{-1/2} I_{(1/4 < \sigma< 2)}
\\
&&\qquad=  \bigl(\sigma^2\bigr)^{-n/2-1/2} \exp \biggl( -
\frac{1}{\sigma^2} \frac{
\sum_{i=1}^{n} (\log(d_i) - \mu)^2}{2} \biggr) I_{(1/4 < \sigma<
2)}
\\
&&\qquad\sim  \operatorname{Inverse \  Gamma} \biggl( \frac{n-1}{2}, \frac{\sum_{i=1}^{n} (\log(d_i) - \mu)^2}{2} \biggr)
I_{(1/4 < \sigma< 2)}.
\end{eqnarray*}

Similarly, the conditional posterior distribution of $\mu$ is
truncated normal,
and so we can also use a Gibbs sampler to update $\mu$. We can see
this as
\begin{eqnarray*}
\pi\bigl(\mu| \sigma^2, N_K, d_i,
y_{ik}, N_k, N\bigr) &\propto & \prod
_{i=1}^{n} e^{-{(\log(d_i) - \mu)^2}/({2 \sigma^2})} I_{(3 < \mu
< 8)} \\
&=& \exp
\biggl( - \frac{ \sum_{i=1}^{n} (\mu- \log(d_i))^2}{2
\sigma^2} \biggr) I_{(3 < \mu< 8)}
\\
&\propto & \exp \Biggl( - \frac{1}{2 \sigma^2} \Biggl( n \mu^2 - 2 \mu
\sum_{i=1}^{n} \log d_i \Biggr)
\Biggr)I_{(3 < \mu< 8)}
\\
&=&  \exp \biggl( - \frac{1}{2 (\sigma^2/n)} \biggl( \mu^2 - 2 \mu
\frac{\sum_{i=1}^{n} \log d_i}{n} \biggr) \biggr) I_{(3 < \mu< 8)}
\\
& \propto & \exp \biggl( - \frac{1}{2 (\sigma^2/n)} \biggl( \mu- \frac
{\sum_{i=1}^{n} \log d_i}{n}
\biggr)^2 \biggr) I_{(3 < \mu< 8)}
\\
&\sim & \operatorname{Normal} \biggl( \frac{\sum_{i=1}^{n} \log d_i}{n}, \frac
{\sigma^2}{n} \biggr)
I_{(3 < \mu< 8)}.
\end{eqnarray*}
Since both $\mu, \sigma$ have uniform priors, if a value is proposed
in the MCMC update outside of the range of the prior, then another
value will be proposed until a value within the range of the prior is proposed.

For $N_K$, the conditional posterior distribution does not have a
closed form,
and so we can use a Metropolis step to update it.
The conditional posterior distribution of $N_K$ is
\begin{eqnarray*}
\pi\bigl(N_K | \mu, \sigma^2, d_i,
y_{ik}, N_k, N\bigr) &\propto & \prod
_{i=1}^{n} \biggl[ \biggl(\frac{N_K}{N}
\biggr)^{y_{iK}} \biggl(1 - \frac{N_K}{N} \biggr)^{d_i - y_{iK}} \biggr]
\cdot\frac{1}{N_K}
\\
&=&  \prod_{i=1}^{n} \biggl[ \biggl(
\frac{N_K}{N-N_K} \biggr)^{y_{iK}} \biggl(1 - \frac{N_K}{N}
\biggr)^{d_i} \biggr] \cdot\frac{1}{N_K},
\end{eqnarray*}
which becomes
\begin{eqnarray*}
&& \ell\bigl(N_K | \mu, \sigma^2, d_i,
y_{ik}, N_k, N\bigr)  \\
&&\qquad \stackrel{c} {=}  \sum
_{i=1}^{n} y_{iK} \log \biggl(
\frac{N_K}{N-N_K} \biggr) + \sum_{i=1}^{n}
d_i \log \biggl(1 - \frac{N_K}{N} \biggr)- \log N_K,
\end{eqnarray*}
in log terms (to maintain numerical stability), where $\stackrel{c}{=}$ denotes equality up to an additive constant.
The proposed value of $N_K$ was rejected if it fell outside the interval
$(\max_{i} y_{ik}, N)$, but this happens rarely.
We used a normal proposal for $N_K$, with the standard deviation being
equal to 2.3 times the residual standard error obtained from regressing
$N_K$ on $\mu$ and $\sigma$ from an initial starting chain
to obtain an appropriate tuning parameter
[\citet{raftery1996implementing}].

The posterior distribution of $d_i$ is
\begin{eqnarray*}
&& \pi\bigl(d_i | \mu, \sigma^2, N_K,
y_{ik}, N_k, N\bigr) \propto\frac{1}{d_i}
e^{-{(\log(d_i) - \mu)^2}/({2 \sigma^2})} \prod_{k=1}^{K}
\pmatrix{d_i \cr  y_{ik}} \biggl(1-\frac{N_k}{N}
\biggr)^{d_i},
\end{eqnarray*}
which results in
\begin{eqnarray*}
&& \ell\bigl(d_i | \mu, \sigma^2, N_K,
y_{ik}, N_k, N\bigr)\\
&&\qquad \stackrel{c} {=}- \log
d_i - \frac{(\log(d_i) - \mu)^2}{2 \sigma^2} + \sum_{k=1}^{K}
\log\pmatrix{d_i \cr  y_{ik}} + \sum
_{k=1}^{K} d_i \log \biggl(1-
\frac
{N_k}{N} \biggr),
\end{eqnarray*}
again in log terms for numerical stability. Just as with $N_K$, we will
reject values of $d_i$ that are below $\max_{k} y_{ik}$.
As before, we used a normal proposal with a tuning parameter
calculated as 2.3 times the residual standard error from a
regression on an initial starting chain.

\subsection{Barrier effects model}\label{A2}

The barrier effects model is defined in Section~\ref{secbarrier.model}.
The posterior distribution is
\begin{eqnarray*}
\label{barrierposterior}
&&\!\! \pi(d_i, \mu, \sigma, m_K,
\rho_k | y_{ik}, N, m_k)
\\
&&\!\!\!\!\!\! \qquad\propto  \prod
_{i=1}^{n} \Biggl[ \frac{1}{d_i \sigma\sqrt{2 \pi}} e^{-{(\log(d_i) - \mu)^2}/({2 \sigma^2})}
\\
&&\!\!\!\!\!\!\hspace*{6pt}\qquad\qquad{}\times
\prod_{k=1}^{K}
\pmatrix{d_i \cr  y_{ik}} \frac{B (m_k
({1}/{\rho_k}-1) + y_{ik}, d_i + (1-m_k)({1}/{\rho
_k}-1)-y_{ik}  )}{B (m_k ({1}/{\rho_k}-1),
(1-m_k)({1}/{\rho_k}-1) )} \Biggr]\\
&&\!\!\!\!\!\!\!\!\qquad\quad{}\times
\frac{1}{m_K} \frac
{1}{1} \frac{1}{5} \frac{1}{7/4}
\end{eqnarray*}
using the beta-binomial distribution, effectively integrating out
$q_{ik}$ and reducing the number of parameters to be sampled. Our MCMC
updates for $\mu, \sigma$ are the same as for the random degree model.

We use a Metropolis step  to update $m_K$, as there is no closed form.
The conditional posterior distribution of $m_K$ is
\begin{eqnarray*}
&& \pi\bigl( m_K | y_{ik}, N, d_i,
\sigma^2, \mu, m_k, \rho_k\bigr)\\
&&\qquad\propto
\prod_{i=1}^{n} \biggl[ \frac{\mathrm{B} (m_K ({1}/{\rho
_K}-1) + y_{iK}, d_i + (1-m_K)({1}/{\rho_K}-1)-y_{iK}
)}{\mathrm{B} (m_K ({1}/{\rho_K}-1), (1-m_K)({1}/{\rho
_K}-1) )}
\biggr] \frac{1}{m_K} ,
\end{eqnarray*}
which becomes
\begin{eqnarray*}
&&\ell\bigl( m_K | y_{ik}, N, d_i,
\sigma^2, \mu, m_k, \rho_k\bigr)\\
&&\qquad \stackrel
{c} {=}\sum_{i=1}^{n} \log\mathrm{B}
\biggl(m_K \biggl(\frac{1}{\rho_K}-1\biggr) + y_{iK},
d_i + (1-m_K) \biggl(\frac{1}{\rho_K}-1
\biggr)-y_{iK} \biggr)
\\
&&\qquad\quad{}- \sum_{i=1}^{n} \log\mathrm{B}
\biggl(m_K \biggl(\frac{1}{\rho_K}-1\biggr), (1-m_K)
\biggl(\frac{1}{\rho_K}-1\biggr) \biggr) - \operatorname{log}(m_K),
\end{eqnarray*}
in log terms (to maintain numerical stability).
The bounds on $m_K$ are $(0, 1)$, as $m_K$ is the proportion of the total
population in subpopulation $K$. We used a normal symmetric reflective
proposal, reflecting values when proposed outside of bounds, as used in
\citet{de2003better}.
For example, if $m_K^{(t)} = 0.9$ and the normal proposal directs
$m_K^{(t+1)} = 1.05$, we would\vspace*{1pt} instead bounce this back such that
$m_K^{(t+1)}$ goes up 0.1, but as that gets to 1, $m_K^{(t+1)}$ then
come down 0.05, resulting in $m_K^{(t+1)} = 0.95$. This distribution is
symmetric, allowing the use of a Metropolis step to update. Just as
with $N_K$ in the random degree model, we will use 2.3 times the
residual standard error from an initial chain as the tuning parameter.

Updating $\rho_k$ will be very similar, with only a difference in the
term for the prior. The conditional posterior for $\rho_k$ is
\begin{eqnarray*}
&& \!\!\!\pi\bigl( \rho_k | y_{ik}, N, d_i,
\sigma^2, \mu, m_k, m_K\bigr) \\
&&\!\!\!\!\!\!\qquad\propto \prod
_{i=1}^{n} \biggl[ \frac{\mathrm{B} (m_K ({1}/{\rho
_K}-1) + y_{iK}, d_i + (1-m_K)({1}/{\rho_K}-1)-y_{iK}
)}{\mathrm{B} (m_K ({1}/{\rho_K}-1), (1-m_K)({1}/{\rho
_K}-1) )} \biggr]
I_{(0 < \rho_k < 1)}\!,
\end{eqnarray*}
which becomes
\begin{eqnarray*}
&&\ell\bigl( \rho_k | y_{ik}, N, d_i,
\sigma^2, \mu, m_k, m_K\bigr) \\
&&\qquad\stackrel{c}{=}\sum_{i=1}^{n} \log\mathrm{B}
\biggl(m_k \biggl(\frac{1}{\rho_k}-1\biggr) + y_{ik},
d_i + (1-m_k) \biggl(\frac{1}{\rho_k}-1
\biggr)-y_{ik} \biggr)
\\
&&\qquad\quad {}-\sum_{i=1}^{n} \log\mathrm{B}
\biggl(m_k \biggl(\frac{1}{\rho_k}-1\biggr), (1-m_k)
\biggl(\frac{1}{\rho_k}-1\biggr) \biggr),
\end{eqnarray*}
in log terms (to maintain numerical stability). Just as with $m_K$,
$\rho_k$ is similarly bounded between 0 and 1. Thus, we have used the
normal symmetric reflective proposal with 2.3 times the residual
standard error as the tuning parameter as well.

Updating $d_i$ can be simplified from the beta functions, as $d_i$ only
appears in one term of the beta function. The posterior for $d_i$\vspace*{-2pt} is
\begin{eqnarray*}
&& \pi\bigl( d_i | y_{ik}, N, \sigma^2, \mu,
m_k, m_K, \rho_k\bigr)\\[-2pt]
&&\qquad \propto
\frac{ e^{-{(\log(d_i) - \mu)^2}/({2 \sigma^2})} }{d_i}\\[-2pt]
&&\qquad\quad {}\times\prod_{k=1}^{K}
\pmatrix{d_i \cr  y_{ik}} \frac{\mathrm{B} (m_k ({1}/{\rho
_k}-1) + y_{ik}, d_i + (1-m_k)({1}/{\rho_k}-1)-y_{ik}
)}{\mathrm{B} (m_k ({1}/{\rho_k}-1), (1-m_k)({1}/{\rho
_k}-1) )}
\\[-2pt]
&&\qquad\propto \frac{e^{-{(\log(d_i) - \mu)^2}/({2 \sigma^2})} }{d_i} \\[-2pt]
&&\qquad\quad{}\times \prod_{k=1}^{K}
\pmatrix{d_i \cr  y_{ik}}\frac{\Gamma(d_i + (1-m_k)
({1}/{\rho_k} - 1) -y_{ik})}{\Gamma(m_k ({1}/{\rho_k} - 1)
+ y_{ik} + d_i + (1-m_k) ({1}/{\rho_k} - 1) -y_{ik})}
\\[-2pt]
&&\qquad\propto \frac{e^{-{(\log(d_i) - \mu)^2}/({2 \sigma^2})} }{d_i} \prod_{k=1}^{K}
\pmatrix{d_i \cr  y_{ik}} \frac{\Gamma(d_i + (1-m_k)
({1}/{\rho_k} - 1) -y_{ik})}{\Gamma( d_i + ({1}/{\rho_k} - 1))},
\end{eqnarray*}
which becomes
\begin{eqnarray*}
&& \ell\bigl( d_i | y_{ik}, N, \sigma^2, \mu,
m_k, m_K, \rho_k\bigr)\\[-2pt]
&&\qquad \stackrel {c} {=}
-\log(d_i) - \frac{(\log(d_i) - \mu)^2}{2 \sigma^2} + \sum_{k=1}^{K}
\log\pmatrix{d_i \cr  y_{ik}}
\\[-2pt]
&&\qquad\quad{}+ \log\Gamma\biggl(d_i + (1-m_k) \biggl(
\frac{1}{\rho_k} - 1\biggr) -y_{ik}\biggr) - \log\Gamma\biggl(
d_i + \biggl(\frac{1}{\rho_k} - 1\biggr)\biggr),
\end{eqnarray*}
in log terms (to maintain numerical stability). As in the random degree
model, $d_i$ must be greater than $\max_{k} y_{ik}$. We again used a
normal proposal with a tuning parameter of 2.3 times the residual
standard error from a regression on an initial starting chain.

\subsection{Transmission bias model}\label{A3}

The transmission bias model is defined in Section~\ref{sectransmission.model}. %
The posterior distribution is
\begin{eqnarray*}
&& \pi\bigl(\mu, \sigma^2, d_i,  N_K,
\tau_{K} | y_{ik}, N_k, N\bigr)\\
 &&\qquad=  \prod
_{i=1}^{n} \frac{1}{d_i \sigma\sqrt{2 \pi}} e^{-{(\log(d_i)
- \mu)^2}/({2 \sigma^2})} \prod
_{k=1}^{K-1} \biggl( \pmatrix{d_i\cr y_{ik}} \biggl(\frac{N_k}{N} \biggr)^{y_{ik}} \biggl(1-
\frac
{N_k}{N} \biggr)^{d_i - y_{ik}} \biggr)
\\
&&\qquad\quad{}\times \prod_{K} \biggl( \pmatrix{d_i \cr
y_{iK}} \biggl(\tau_{K} \frac
{N_{K}}{N}
\biggr)^{y_{iK}} \biggl(1-\tau_{K}\frac{N_{K}}{N}
\biggr)^{d_i - y_{iK}} \biggr)\\
&&\hspace*{25pt}\qquad\quad{}\times \frac{\tau_{K}^{\eta_K ({1}/{\nu_K}
- 1) - 1} (1-\tau_{K})^{(1 - \eta_K) ({1}/{\nu_K} - 1) -
1}}{B(\eta_K ({1}/{\nu_K} - 1), (1 - \eta_K) ({1}/{\nu
_K} - 1))} \frac{1}{N_K}
\frac{1}{5} \frac{1}{7/4}.
\end{eqnarray*}

Since $\tau_K$ and $N_K$ are not very clearly identifiable and tend
to be highly correlated a posteriori,
and were mirroring each other in early MCMC chains,
we reparametrized the model using
\begin{eqnarray*}
&& w_K = N_K \tau_K, \qquad z_K =
\frac{N_K}{\tau_K}.
\end{eqnarray*}
To compute the Jacobian, we have
\begin{eqnarray*}
&& N_K = \sqrt{w_K z_K}, \qquad
\tau_K = \sqrt{\frac{w_K}{z_K}}.
\end{eqnarray*}
Thus, the determinant of the Jacobian is
\begin{eqnarray*}
\left|\frac{\partial(N_K, \tau_K)}{\partial(w_K, z_K)}\right|
& =&
\left|\matrix{\displaystyle
\frac{\partial}{\partial w_K} \sqrt{w_K
z_K} & \displaystyle\frac
{\partial}{\partial z_K} \sqrt{w_K z_K}
\vspace*{3pt}\cr
\displaystyle\frac{\partial}{\partial
w_K} \sqrt{\frac{w_K}{z_K}} & \displaystyle\frac{\partial}{\partial z_K} \sqrt {
\frac{w_K}{z_K}}}\right|
\\
&= &
\left|\matrix{\displaystyle
\frac{1}{2} \sqrt{\frac{z_K}{w_K}} &
\displaystyle\frac{1}{2} \sqrt{\frac{w_K}{z_K}}
\vspace*{3pt}\cr
\displaystyle\frac{1}{2} \sqrt{\frac{1}{w_K z_K}} & \displaystyle-\frac{1}{2} \sqrt
{\frac{w_K}{z_K^3}} }\right|
\\
&= & -\frac{1}{4 z_K} - \frac{1}{4 z_K} = -\frac{1}{2 z_K}.
\end{eqnarray*}

The reparameterized posterior, in terms of $w_K, z_K$, is thus
\begin{eqnarray*}
&& \!\!\pi\bigl(\mu, \sigma^2, d_i, w_K,
z_K | y_{ik}, N_k, N, \eta_K,
\nu_K\bigr) \\
&&\!\!\!\!\!\!\qquad= \prod_{i=1}^{n}
\frac{1}{d_i \sigma\sqrt{2 \pi}} e^{-{(\log(d_i) - \mu)^2}/({2 \sigma^2})} \prod_{k=1}^{K-1}
\biggl( \pmatrix{d_i \cr y_{ik}} \biggl(\frac{N_k}{N}
\biggr)^{y_{ik}} \biggl(1-\frac{N_k}{N} \biggr)^{d_i - y_{ik}} \biggr)
\\
&&\!\!\!\!\!\!\qquad\quad{}\times \prod_{K} \biggl( \pmatrix{d_i\cr y_{iK}} \biggl( \frac{w_K}{N} \biggr)^{y_{iK}} \biggl(1-
\frac{w_K}{N} \biggr)^{d_i - y_{iK}} \biggr) \frac{1}{\sqrt{w_K z_K}}
\\
&&\!\!\!\!\!\!\hspace*{24pt}\qquad\quad{}\times \frac{\sqrt{{w_K}/{z_K}}^{\eta_K ({1}/{\nu_K} - 1) -
1} (1-\sqrt{{w_K}/{z_K}})^{(1 - \eta_K) ({1}/{\nu_K} - 1)
- 1}}{B(\eta_K ({1}/{\nu_K} - 1), (1 - \eta_K) ({1}/{\nu
_K} - 1))} \frac{1}{5} \frac{1}{7/4} \frac{1}{2 z_K}\!.
\end{eqnarray*}

We can update $\mu, \sigma$ as in the previous models.

The conditional posterior of $d_i$ is
\begin{eqnarray*}
&& \pi\bigl(d_i | \mu, \sigma^2, w_K,
z_K, y_{ik}, N_k, N, \eta_K,
\nu_K\bigr) \\
&&\qquad\propto\frac{1}{d_i} e^{-{(\log(d_i) - \mu)^2}/({2 \sigma^2})} \prod
_{k=1}^{K} \pmatrix{d_i\cr y_{ik}} \biggl(1-\frac{w_k}{N} \biggr)^{d_i},
\end{eqnarray*}
which results in
\begin{eqnarray*}
&& \ell\bigl(d_i | \mu, \sigma^2, N_K,
\tau_K, y_{ik}, N_k, N, \eta_K,
\nu _K\bigr) \\
&&\qquad\stackrel{c} {=}- \log d_i -
\frac{(\log(d_i) - \mu)^2}{2 \sigma
^2} + \sum_{k=1}^{K} \log
\pmatrix{d_i \cr  y_{ik}} + \sum_{k=1}^{K}
d_i \log \biggl(1-\frac{w_k}{N} \biggr) ,
\end{eqnarray*}
in log terms for numerical stability. Note that this equation calls for
a $w_k$ for $k$ from the known subpopulations as well. Since $w_K =
\tau_K N_K$ and we are assuming $\tau_k = 1$ for $k$ known (no
transmission bias present in known subpopulations), we have $w_k = N_k$
in the known subpopulations. As before, we used a normal proposal for
$d_i$, keeping the old value when we propose a value less than $\max_k
y_{ik}$, and used~2.3 times the residual standard error for the tuning
parameter.

Now, instead of updating $N_K, \tau_K$, we can update $w_K, z_K$ as
given below.
The conditional posterior for $w_K$ is
\begin{eqnarray*}
&& \pi\bigl(w_K | \mu, \sigma^2, d_i,
z_K y_{ik}, N_k, N, \eta_K,
\nu_K\bigr)\\
 &&\qquad\propto \prod_{i=1}^{n}
\biggl[ \biggl( \frac{w_K}{N} \biggr)^{y_{iK}} \biggl(1-
\frac{w_K}{N} \biggr)^{d_i - y_{iK}} \biggr] \frac{1}{\sqrt{w_K}}
\sqrt{w_K}^{\eta_K ({1}/{\nu_K} - 1) -
1}
\\
&&\qquad\quad\hspace*{12pt}{}\times \biggl(1-\sqrt{\frac{w_K}{z_K}} \biggr)^{(1 - \eta_K) ({1}/{\nu_K} - 1) - 1} \frac{1}{2 z_K}
\\
&&\qquad= \prod_{i=1}^{n} \biggl[ \biggl(
\frac{w_K}{N - w_K} \biggr)^{y_{iK}} \biggl(1-\frac{w_K}{N}
\biggr)^{d_i} \biggr] \sqrt {w_K}^{\eta_K ({1}/{\nu_K} - 1) - 2}
\\
&&\qquad\quad\hspace*{12pt}{}\times \biggl(1-\sqrt{\frac{w_K}{z_K}} \biggr)^{(1 - \eta_K) ({1}/{\nu_K} - 1) - 1} \frac{1}{2 z_K}.
\end{eqnarray*}
This results in
\begin{eqnarray*}
&& \ell\bigl(w_K | \mu, \sigma^2, d_i,
z_K y_{ik}, N_k, N, \eta_K,
\nu_K\bigr) \\
&&\qquad\stackrel{c} {=} \sum_{i=1}^n
y_{iK} \log \biggl(\frac{w_K}{N - w_K} \biggr) + \sum
_{i=1}^{n} d_i \biggl( 1 -
\frac{w_K}{N} \biggr)
\\
&&\qquad\quad{}+ \frac{\eta_K ({1}/{\nu_K} - 1) - 2}{2} \log w_K + \biggl((1 - \eta_K)
\biggl(\frac{1}{\nu_K} - 1\biggr) - 1\biggr)
\\
&&\qquad\quad{}\times\log \biggl(1 - \sqrt{
\frac
{w_K}{z_K}} \biggr)- \log2z_K,
\end{eqnarray*}
in log terms for numerical\vadjust{\goodbreak} stability.

The posterior for $z_K$ is
\begin{eqnarray*}
&& \pi\bigl(z_K | \mu, \sigma^2, d_i,
w_K y_{ik}, N_k, N, \eta_K,
\nu_K\bigr)\\
 &&\qquad \propto  \frac{1}{\sqrt{z_K}} \sqrt{\frac{1}{z_K}}^{\eta_K ({1}/{\nu_K} - 1) - 1}
\biggl(1-\sqrt{\frac{w_K}{z_K}} \biggr)^{(1 -
\eta_K) ({1}/{\nu_K} - 1) - 1} \frac{1}{2 z_K}
\\
&&\qquad= {z_K}^{-{ \eta_K ({1}/{\nu_K} - 1)}/{2}} \biggl(1-\sqrt {\frac{w_K}{z_K}}
\biggr)^{(1 - \eta_K) ({1}/{\nu_K} - 1) -
1} \frac{1}{2 z_K},
\end{eqnarray*}
which results in
\begin{eqnarray*}
&& \ell\bigl(z_K | \mu, \sigma^2, d_i,
w_K y_{ik}, N_k, N, \eta_K,
\nu_K\bigr)\\
 &&\qquad\stackrel{c} {=} -\frac{\eta_K ({1}/{\nu_K} - 1)}{2} \log
z_K + \biggl((1 - \eta_K) \biggl(\frac{1}{\nu_K} - 1
\biggr) - 1\biggr) \log \biggl(1-\sqrt {\frac{w_K}{z_K}} \biggr)
 \\
 &&\qquad\quad{}- \log2z_K,
\end{eqnarray*}
in log terms for numerical stability. Both $w_K$ and $z_K$ must be
positive and $w_K$ must be larger than $z_K$. The parameter $w_K$
cannot be larger than the total population, but $z_K$ does not have a
clear upper bound, except that $N_K = \sqrt{w_K z_K}$ must be less
than the total population. All relevant bounds are included, rejecting
proposed values of $w_K$ or $z_K$ if the they do not fall within the
bounds. As for other parameters, for the tuning parameter, we used 2.3
times the residual standard error obtained by a regression from a small
initial chain.

\subsection{Barrier transmission combined model}\label{A4}

The combined barrier effects and transmission bias model is defined
in Section~\ref{seccombined.model}.
The posterior for this model is
\begin{eqnarray*}
\label{posteriorcombined}
&&\pi(\mu, \sigma, d_i, m_k, \rho_k,
q_{ik}, \tau_{k} | y_{ik}, m_k, N,
\eta_K, \nu_K) \\
&&\qquad= \prod_{i=1}^{n}
\frac{1}{d_i \sigma\sqrt
{2 \pi}} e^{-{(\log(d_i) - \mu)^2}/({2 \sigma^2})}\\
&&\qquad\quad{}\times \prod_{k=1}^{K}
\biggl( \pmatrix{d_i \cr  y_{ik}} (\tau_k
q_{ik} )^{y_{ik}} (1- \tau_k q_{ik}
)^{d_i - y_{ik}}
\\
&&\qquad\qquad\hspace*{23pt}{}\times\frac{q_{ik}^{m_k ({1}/{\rho_k} - 1) -1}
(1-q_{ik})^{(1-m_k) ({1}/{\rho_k} - 1) -1}}{B(m_k ({1}/{\rho
_k} - 1), (1-m_k)({1}/{\rho_k} - 1)} \biggr)\\
 &&\qquad\quad{}\times\prod_{K}
\frac
{\tau_K^{\eta_K ({1}/{\nu_K} - 1) - 1} (1-\tau_K)^{(1-\eta_K)
({1}/{\nu_K} - 1) -1}}{B(\eta_K ({1}/{\nu_K} - 1), (1-\eta
_K) ({1}/{\nu_K} - 1))}
 \frac{1}{m_K} \frac{1}{1} \frac{1}{5} \frac{1}{7/4}.
\end{eqnarray*}
Note that $\eta_K$ and $\nu_K$ in the distribution of $\tau_K$ would be
fixed based on external information. We cannot use the beta-binomial
distribution to integrate out $q_{ik}$ due to the $\tau_K$ in the
model; thus, we must sample $q_{ik}$ as well, significantly increasing
the number of parameters in the model.

The updates for $\mu, \sigma$ are as in the previous models.

We can update $m_K$ using a Metropolis step just as in the other models.
The conditional posterior for $m_K$ is
\begin{eqnarray*}
&& \pi( m_K | y_{ik}, N, d_i, \sigma, \mu,
m_k, \rho_k, q_{ik}, \tau _K,
\eta_K, \nu_K) \\
&&\qquad\propto\prod
_{i=1}^{n} \biggl(\frac{q_{iK}^{m_K ({1}/{\rho_K} - 1) -1} (1-q_{iK})^{(1-m_K) ({1}/{\rho_K} - 1) -1}}{\mathrm{B}(m_K ({1}/{\rho_K} - 1),
(1-m_K)({1}/{\rho_K} - 1)} \biggr)
\frac{1}{m_K},
\end{eqnarray*}
which becomes
\begin{eqnarray*}
&& \ell( m_K | y_{ik}, N, d_i, \sigma, \mu,
m_k, \rho_k, q_{ik}, \tau _K,
\eta_K, \nu_K) \\
&&\qquad\stackrel{c} {=}\sum
_{i=1}^{n} \biggl[\biggl(m_K \biggl(
\frac{1}{\rho_K}-1\biggr) -1\biggr) \log q_{iK}
 \\
 &&\hspace*{9pt}\qquad\qquad{}+ \biggl((1-m_K) \biggl(\frac{1}{\rho_K}-1\biggr) -1
\biggr) \log(1-q_{iK} )
\\
 &&\hspace*{9pt}\qquad\qquad{}- \log\mathrm{B} \biggl(m_K \biggl(\frac{1}{\rho_K}-1
\biggr), (1-m_K) \biggl(\frac{1}{\rho_K}-1\biggr) \biggr) \biggr] -
\operatorname{log}(m_K),
\end{eqnarray*}
in log terms (to maintain numerical stability).

Updating $\rho_k$ is similar, just with a different prior.
The conditional posterior for $\rho_k$ is
\begin{eqnarray*}
&& \pi(\rho_k | y_{ik}, N, d_i, \sigma, \mu,
m_k, m_K, q_{ik}, \tau _K,
\eta_K, \nu_K) \\
&&\qquad\propto\prod
_{i=1}^{n} \biggl(\frac{q_{ik}^{m_k ({1}/{\rho_k} - 1) -1} (1-q_{ik})^{(1-m_k) ({1}/{\rho_k} - 1) -1}}{\mathrm{B}(m_k ({1}/{\rho_k} - 1),
(1-m_k)({1}/{\rho_k} - 1)} \biggr),
\end{eqnarray*}
which becomes
\begin{eqnarray*}
&& \ell( \rho_K | y_{ik}, N, d_i, \sigma, \mu,
m_k, m_K, q_{ik}, \tau _K,
\eta_K, \nu_K) \\
&&\qquad \stackrel{c}{=} \sum
_{i=1}^{n} \biggl[\biggl(m_k \biggl(
\frac{1}{\rho_k}-1\biggr) -1\biggr) \log q_{ik}
\\
&&\qquad\qquad\hspace*{9pt}{}+ \biggl((1-m_k) \biggl(\frac{1}{\rho_k}-1\biggr) -1
\biggr) \log(1-q_{ik} )
\\
&&\qquad\qquad\hspace*{9pt}{}- \log\mathrm{B} \biggl(m_k \biggl(\frac{1}{\rho_k}-1
\biggr), (1-m_k) \biggl(\frac{1}{\rho_k}-1\biggr) \biggr) \biggr],
\end{eqnarray*}
in log terms (to maintain numerical stability).
Note that $\tau_k = 1$ for $k$ a known population.

To update $q_{ik}$, the conditional posterior is
\begin{eqnarray*}
&& \pi( q_{ik} | y_{ik}, N, d_i, \sigma, \mu,
m_k, m_K, \rho_k, \tau _K,
\eta_K, \nu_K) \\
&&\qquad\propto{q_{ik}}^{y_{ik}}
(1- \tau_k q_{ik} )^{d_i - y_{ik}} q_{ik}^{m_k ({1}/{\rho_k} - 1)
-1}
(1-q_{ik})^{(1-m_k) ({1}/{\rho_k} - 1) -1}
\\
&&\qquad\propto q_{ik}^{y_{ik} + m_k ({1}/{\rho_k} - 1) -1} (1-\tau_k
q_{ik})^{d_i - y_{ik}}(1-q_{ik})^{(1-m_k) ({1}/{\rho_k} - 1) -1},
\end{eqnarray*}
which becomes\vspace*{-1pt}
\begin{eqnarray*}
&& \ell( q_{ik} | y_{ik}, N, d_i, \sigma, \mu,
m_k, m_K, \rho_k, \tau _K,
\eta_K, \nu_K)\\[-1pt]
&&\qquad \stackrel{c} {=} \biggl(y_{ik}
+ m_k \biggl(\frac{1}{\rho
_k} - 1\biggr) -1\biggr) \log
q_{ik} + (d_i - y_{ik}) \log(1-
\tau_k q_{ik})
\\[-1pt]
&&\qquad\quad {}+ \biggl((1-m_k) \biggl(\frac{1}{\rho_k} - 1\biggr) -1\biggr)
\log(1-q_{ik}),
\end{eqnarray*}
in log terms (to maintain numerical stability).
Again, note that $\tau_k = 1$ when $k$ represents a known population.

To update $\tau_K$, the posterior is
\begin{eqnarray*}
&& \pi(\tau_K | y_{ik}, N, d_i, \mu, \sigma,
m_k, m_K, \rho_k, q_{iK},
\eta_K, \nu_K)\\
&&\qquad \propto  \prod
_{i=1}^{n} \tau_{K}^{y_{iK}} (1-
\tau_{K} q_{iK} )^{d_i - y_{iK}} \tau_{K}^{\eta_K
({1}/{\nu_K} - 1) - 1}
(1-\tau_{K})^{(1-\eta_K) ({1}/{\nu_K} - 1) - 1}
\\
&&\qquad\propto  \prod_{i=1}^{n}
\tau_{K}^{y_{ik} + \eta_K ({1}/{\nu_K} - 1) - 1} (1-\tau_K q_{iK})^{d_i - y_{iK}}
(1-\tau_K)^{(1-\eta
_K) ({1}/{\nu_K} - 1) - 1},
\end{eqnarray*}
which results in\vspace*{-1pt}
\begin{eqnarray*}
&&\ell(\tau_K | y_{ik}, N, d_i, \mu, \sigma,
m_k, m_K, \rho_k, q_{iK},
\eta_K, \nu_K)\\[-1pt]
 &&\qquad\stackrel{c} {=}\sum
_{i=1}^{n} \biggl(y_{iK} +\eta
_K \biggl(\frac{1}{\nu_K} - 1\biggr) - 1\biggr)\log
\tau_{K}
\\[-1pt]
&&\qquad\quad{}+ \sum_{i=1}^{n} (d_i -
y_{iK}) \log (1-\tau_{K} q_{iK} )
 \\[-1pt]
 &&\qquad\quad{}+n \biggl((1-\eta_K) \biggl(\frac{1}{\nu_K} - 1\biggr) - 1
\biggr) \log(1-\tau_K),
\end{eqnarray*}
in log terms (to maintain numerical stability).\vadjust{\goodbreak}

All of $m_K$, $\rho_k$, $p_{ik}$ and $\tau_K$ are constrained to lie
between 0 and 1. We again used the normal symmetric reflective
proposal, reflecting values outside of the allowed range. We also again
used 2.3 times the residual standard error from an initial chain for
the tuning parameter.

The posterior for $d_i$ is\vspace*{-1pt}
\begin{eqnarray*}
&& \pi( d_i | y_{ik}, N, \mu, \sigma, m_k,
m_K, \rho_k, q_{iK}, \tau _K,
\eta_K, \nu_K)\\[-1pt]
&&\qquad \propto  \frac{1}{d_i}
e^{-{(\log(d_i) -
\mu)^2}/({2 \sigma^2})} \prod_{k=1}^{K}
\pmatrix{d_i \cr  y_{ik}} (1 - \tau _k
q_{ik})^{d_i},\vspace*{-1pt}
\end{eqnarray*}
which becomes\vspace*{-1pt}
\begin{eqnarray*}
&&\ell( d_i | y_{ik}, N, \mu, \sigma, m_k,
m_K, \rho_k, q_{iK}, \tau _K,
\eta_K, \nu_K) \\[-1pt]
&&\qquad\stackrel{c} {=}  -\log(d_i)
- \frac{(\log
(d_i) - \mu)^2}{2 \sigma^2} + \sum_{k=1}^{K}
\biggl[ \log\pmatrix{d_i \cr  y_{ik}}
+ d_i \log(1-\tau_k q_{ik})
\biggr],\vspace*{-1pt}
\end{eqnarray*}
in log terms (to maintain numerical stability).
For $d_i$, we again used a normal proposal with a tuning parameter
of 2.3 times the residual standard error.\vspace*{-1pt}
\end{appendix}


%
%




\printaddresses
\end{document}